\documentclass[prd,aps,nofootinbib,10pt]{revtex4}
\usepackage{amsmath,graphicx,epsfig,amssymb,dsfont,mathtools}
\usepackage[usenames]{color}
\usepackage{ulem} 
\usepackage{bigstrut}

\begin{document}
\title{$a_1(1260), a_1(1420)$ and the production in heavy meson decays }
\author{ Wei Wang$^{1,2}$~\footnote{Email:{wei.wang@sjtu.edu.cn}}, and Zhen-Xing Zhao$^{1}$~\footnote{Email:{star\_0027@sjtu.edu.cn}}}

\affiliation{
$^1$ INPAC, Shanghai Key Laboratory for Particle Physics and Cosmology, Department of Physics and Astronomy, Shanghai Jiao-Tong University, Shanghai, 200240,   China\\
$^2$
State Key Laboratory of Theoretical Physics, Institute of Theoretical Physics, Chinese Academy of Sciences, Beijing 100190, China}

\begin{abstract}
The   $a_1(1420)$   with $I^G(J^{PC})= 1^-(1^{++})$  observed in the $\pi^+ f_0(980)$ final state  in the $\pi^-p\to \pi^+\pi^-\pi^- p$   process  by  the COMPASS collaboration seems unlikely to be an ordinary  $\bar qq$ mesonic state. Available theoretical explanations include tetraquark or rescattering effects due to $a_1(1260)$ decays.  If the $a_1(1420)$ were induced by the rescattering,  its production rates are completely determined by those of the $a_1(1260)$.   In this work, we propose to explore the ratios of branching fractions of  heavy meson weak decays into  the $a_1(1420)$ and $a_1(1260)$, and testing the universality of these ratios  would be a straightforward way to validate/invalidate the rescattering explanation. The decay modes include in the charm sector  the $D^0\to  a_1^-\ell^+\nu$ and  $D^0\to \pi^\pm a_1^\mp$, and  in the bottom sector   $\overline B^0\to  a_1^+ \ell^- \bar\nu$, $B\to D  a_1, \pi^\pm a_1^\mp$,   $B_c\to J/\psi a_1$ and $\Lambda_b\to \Lambda_c a_1$.  We calculate the branching ratios for   various decay modes into the $a_1(1260)$. The numerical  results indicate that there is a promising prospect to  study  these decays on experiments including BES-III, LHCb, Babar, Belle and CLEO-c,  the forthcoming  Super-KEKB factory and the under-design Circular Electron-Positron Collider. Experimental analyses in future will  lead to a deeper understanding of the nature of the $a_1(1420)$.
\end{abstract}

\maketitle


\section{Introduction}

Since Gell-Mann proposed the concept of quarks in 1964~\cite{GellMann:1964nj},    quark model has achieved  indisputable sucesses:  most of the established  mesons and baryons on experimental side can be well accommodated in  the predicted scheme~\cite{Agashe:2014kda}.
However, recently  there have been experimental  
observations of resonance-like structures with
quantum numbers  hardly to be placed  in  the quark-antiquark or three-quark
schemes~\cite{Choi:2003ue,Choi:2007wga,Belle:2011aa,Ablikim:2013mio,Liu:2013dau,Xiao:2013iha,Aaij:2015tga}. This leads to the suspect that  the hadron spectrum is much richer than the simple quark model~\cite{Brambilla:2010cs}.

\begin{figure}\begin{center}
\includegraphics[scale=0.6]{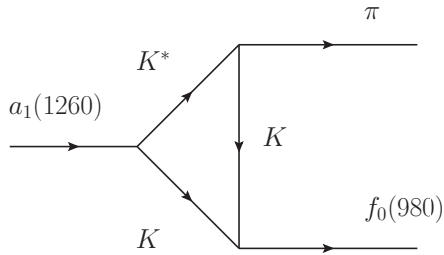}
\caption{Illustration of the $a_1^\pm(1260)\to \pi^\pm f_0(980)$  } \label{fig:rescattering}
\end{center}
\end{figure}

Recently the COMPASS collaboration~\cite{Adolph:2015pws,Ketzer:2014raa} has reported the observation of a light resonance-like state with quantum numbers $I^G (J^{PC}) = 1^- (1^{++})$ in the  $P$-wave
$f_0(980) \pi$ final state with $f_0(980) \to \pi^+\pi^-$.    The signal was also confirmed by the VES experiment
\cite{Khokhlov:2014nha} in the $\pi^-\pi^0\pi^0$ final state. The new state
was tentatively called $a_1(1420)$ with the  mass $m_{a_1} \approx 1.42\,$GeV
and width $\Gamma_{a_1}\approx 0.14\,$GeV. The interpretation of this
state as a new $\bar qq$ meson is
challenging, since  it could hardly  be accommodated as  the  radial excitation of the $a_1(1260)$ which is
expected to have a mass above $1650\,$MeV.  Therefore, this state has been interpreted 
as a tetra-quark~\cite{Chen:2015fwa} or some
dynamical effects arising  from final state interactions~\cite{Ketzer:2015tqa,Liu:2015taa}.
An illustration of the rescattering  mechanism is shown in Fig.~\ref{fig:rescattering}.

The deciphering  of the internal structure  of the $a_1(1420)$  can  proceed
not only through the detailed  analysis of the  pole position, but also through the decay and  production
characters.  In this work, we propose that  semileptonic and nonleptonic heavy meson decays  can be used to examine the rescattering interpretation. In particular, an intriguing property in the rescattering picture  is that  the production rates of $a_1(1420)$ are completely  determined by those of the $a_1(1260)$.  In this case, the ratios
\begin{eqnarray}
 R(B\to a_1 X) = \frac{{\cal B}(B\to a_1^\pm(1420) X){\cal B}( a_1^\pm(1420)\to f_0(980)\pi^\pm)}{{\cal B}(B\to a_1^\pm(1260) X){\cal B}( a_1^\pm(1260)\to 2\pi^\pm\pi^\mp)} ,\\
 R(D\to a_1 Y) = \frac{{\cal B}(D\to a_1^\pm(1420) Y){\cal B}( a_1^\pm(1420)\to f_0(980)\pi^\pm)}{{\cal B}(D\to a_1^\pm(1260) Y){\cal B}( a_1^\pm(1260)\to 2\pi^\pm\pi^\mp)},
\end{eqnarray}
would be insensitive to the production mechanism, and thus would be  a constant.  The value for the ratios  is estimated to be at percent level in Ref.~\cite{Ketzer:2015tqa}.  Testing the universality of these ratios will be a straightforward way to validate/invalidate the rescattering interpretation.

In the above equations,  the  $X,Y$ correspond to  certain leptonic/hadronic final states, and more explicitly  we suggest to study in the charm sector  the $D\to a_1 \ell^+\nu$ and  $D^0\to \pi^\pm a_1^\mp$, and  in the bottom sector  the $B\to a_1 \ell^-\bar\nu$,   $B\to Da_1, \pi^\pm a_1^\mp$,  the $B_c\to J/\psi a_1$ and $\Lambda_b\to \Lambda_c  a_1$ decays. In the following of this work, we will provide the theoretical calculation of branching ratios for the $B$ or $D$ decays into $a_1(1260)$.  In addition, the $B/D$ decays into the $a_1(1260)$ are also of great interest since they are helpful to understand the  dynamics in the decays into an axial-vector meson. Some theoretical  studies  can be found  in the literature~\cite{Cheng:2003sm,Yang:2008xw,Li:2009tx,Verma:2011yw,YanJun:2011rn,Ebert:2011ry,Wang:2008bw,Mehraban:2014hxa,Zhang:2012ev,Liu:2012jb,Cheng:2010vk}.

The rest of this paper is organized as follows. In Sec.~\ref{sec:Ba1}, we will concentrate on  the $B\to a_1(1260)$ decays, including the transition form factors,  semileptonic and non-leptonic  decay modes. We will subsequently  discuss the production of the $a_1(1260)$ in semileptonic and nonleptonic $D/D_s$ decays in Sec.~\ref{sec:Da1}. The last section contains our summary. 

\section{$B$ decays into $a_1$}
\label{sec:Ba1}

\subsection{Form factors}

\begin{figure}\begin{center}
\includegraphics[scale=0.6]{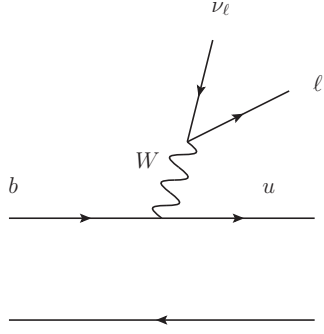}
\caption{Feynman diagram  for the semileptonic  $B\to a_1\ell\bar\nu$ decay.  } \label{fig:feyman_form}
\end{center}
\end{figure}

Unless specified  in the following, we will use the abbreviation $a_1$ to denote the $a_1(1260)$ for simplicity.
The Feynman diagram for semileptonic $\bar B^0\to a_1^+\ell^-\bar\nu$ decays   is given in Fig.~\ref{fig:feyman_form}.
After integrating out the off-shell $W$ boson, one obtains the
effective Hamiltonian
 \begin{eqnarray}
 {\cal H}_{\rm eff} =\frac{G_F}{\sqrt{2}}V_{ub}[\bar
 u\gamma_{\mu}(1-\gamma_5)b] [\bar  {\ell}\gamma^{\mu}(1-\gamma_5)\nu_{\ell}].
 \end{eqnarray}
Here the $V_{ub}$ is the CKM matrix element, and $G_F$ is the Fermi constant.

Hadronic effects are parametrized in terms of the $B\to a_1$  form factors:
 \begin{eqnarray}
  \langle a_1(p_{a_1},\epsilon)|\bar u\gamma^{\mu}\gamma_5 b|\overline B(p_B)\rangle
   &=&-\frac{2iA(q^2)}{m_B-m_{a_1}}\epsilon^{\mu\nu\rho\sigma}
     \epsilon^*_{\nu}p_{B\rho}p_{{a_1}\sigma}, \nonumber\\
  \langle {a_1}(p_{a_1},\epsilon)|\bar u\gamma^{\mu}b|
  \overline B(p_B)\rangle
   &=&-2m_{a_1} V_0(q^2)\frac{\epsilon^*\cdot q}{q^2}q^{\mu}
    -(m_B-m_{a_1})V_1(q^2)\left[\epsilon^{*\mu}
    -\frac{\epsilon^*\cdot q}{q^2}q^{\mu} \right] \nonumber\\
    &&+V_2(q^2)\frac{\epsilon^*\cdot q}{m_B-m_{a_1}}
     \left[ (p_B+p_{a_1})^{\mu}-\frac{m_B^2-m_{a_1}^2}{q^2}q^{\mu} \right],
 \end{eqnarray}
 with $q=p_{B}-p_{a_1}$, and $\epsilon^{0123}=+1$.

The $B\to a_1(1260)$ form factors have been studied  in the covariant light-front quark model (LFQM)~\cite{Cheng:2003sm}, light-cone sum rules (LCSR)~\cite{Yang:2008xw} and perturbative QCD approach (PQCD)~\cite{Li:2009tx}. The corresponding  results  are collected   in Table~\ref{Tab:formFactor}.
In order to access the form factors  in
the full kinematics region, one has  adopted  the dipole parametrization~\cite{Cheng:2003sm,Yang:2008xw,Li:2009tx}:
 \begin{eqnarray}
 F(q^2)=\frac{F(0)}{1-a(q^2/m_B^2)+b(q^2/m_B^2)^2}\;.
 \end {eqnarray}
In the PQCD approach~\cite{Li:2009tx}, the form factor $V_2$ is parametrized  as
 \begin{eqnarray}
 V_2(q^2)&=&\frac{1}{\eta}\big[(1-r_{a_1})^2V_1(q^2)-2r_{a_1}(1-r_{a_1})V_0(q^2)\big].
 \end{eqnarray}
 with $\eta= 1-q^2/m_B^2$, and $r_{a_1}=m_{a_1}/m_{B}$.

\begin{table}
\caption{Results for the $B\to a_1(1260)$ form factors calculated in the covariant light-front quark model (LFQM)~\cite{Cheng:2003sm}, light-cone sum rules (LCSR)~\cite{Yang:2008xw} and perturbative QCD approach (PQCD)~\cite{Li:2009tx}.}
 \label{Tab:formFactor}
 \begin{center}
 \begin{tabular}{|cc|c|c|c|c|c|c|c|c|c|c|c|c|c|c|cc|c|c|c|c|c|c}
\hline \hline
         & $F(0)$     & LFQM&         LCSR  & PQCD    & $a$     & LFQM&         LCSR   & PQCD   & $b$     & LFQM&         LCSR   & PQCD        \\
 \hline
\hline
\ \ \    & $A$                &$0.25$       &$0.48\pm 0.09$     &$0.26_{-0.05-0.01-0.03}^{+0.06+0.00+0.03}$    & $A$                &$1.51$       &$1.64$     &$1.72_{-0.05}^{+0.05}$    & $A$                &$0.64$       &$0.986$     &$0.66_{-0.06}^{+0.07}$       \\
\hline   & $V_0$        &$0.13$       &$0.30\pm 0.05$   &$0.34_{-0.07-0.02-0.08}^{+0.07+0.01+0.08}$         & $V_0$ &$1.71$       &$1.77$     &$1.73_{-0.06}^{+0.05}$    & $V_0$                &$1.23$       &$0.926$     &$0.66_{-0.08}^{+0.06}$      \\
\hline   & $V_1$         &$0.37$       &$0.37\pm 0.07$    &$0.43_{-0.09-0.01-0.05}^{+0.10+0.01+0.05}$         & $V_1$ &$0.29$       &$0.645$     &$0.75_{-0.05}^{+0.05}$    & $V_1$                &$0.14$       &$0.250$     &$-0.12_{-0.02}^{+0.05}$      \\
\hline   & $V_2$                                              &$0.18$       &$0.42\pm 0.08$       &$0.13_{-0.03-0.01-0.00}^{+0.03+0.00+0.00}$         & $V_2$ &$1.14$       &$1.48$     &$--$    & $V_2$                &$0.49$       &$1.00$     &$--$      \\
\hline
\end{tabular}
\end{center}
\end{table}

\subsection{Semileptonic $\overline B^0\to a_1^+(1260)\ell^-\bar\nu_{\ell}$ Decays}

Decay amplitudes for the $\overline B^0\to a_1^+(1260)\ell^-\bar\nu_{\ell}$  can be divided into  hadronic and leptonic sectors. Each of them are expressed in terms of the  Lorentz invariant helicity amplitudes.
The hadronic amplitude is  obtained by  evaluating the matrix element:
\begin{eqnarray}
 i {\cal A}^1_{\lambda}  =   \sqrt{ N_{a_1}}   \frac{iG_{F}}{\sqrt 2} V_{ub} \epsilon_{\mu}^*(h) \langle a_1 |\bar u \gamma^\mu(1-\gamma_5) b |\overline B\rangle,
\end{eqnarray}
with  $\hat m_l= m_l/\sqrt{q^2}$,  $\beta_l=(1-m_l^2/q^2)$ and
\begin{eqnarray}
N_{a_1}= \frac{8}{3}  \frac{\sqrt{\lambda} q^2\beta_l^2 }{ 256 \pi^3m_{B}^3 },\;\;\;
\lambda\equiv \lambda(m_B^2,m_{a_1}^2,q^2)=(m_{B}^2+m_{a_1}^2-q^2)^2-4m_B^2m_{a_1}^2.
\end{eqnarray}
In the above, $\epsilon_\mu(h)$ with $h= 0, \pm, t$ is an auxiliary polarisation vector for the lepton pair system.
The  polarised decay amplitudes are  evaluated as
\begin{eqnarray}
 i{\cal A}^1_0&=&- \sqrt{ N_{a_1}}\frac{  N_1  i}{2m_{a_1}\sqrt {q^2}}\left[   (m_{B}^2-m_{a_1}^2-q^2)(m_{B}-m_{a_1})V_1(q^2)
 -\frac{\lambda}{m_{B}-m_{a_1}}V_2(q^2)\right], \nonumber\\
 i{\cal A}^1_\pm
 &=&   \sqrt{ N_{a_1}}N_1  i \left[  (m_{B}-m_{a_1})V_1(q^2)\mp \frac{\sqrt \lambda}{m_{B}-m_{a_1}}A(q^2) \right],\\
 i{\cal A}^1_t&=&-  i \sqrt{ N_{a_1}} N_{1}  \frac{\sqrt \lambda}{\sqrt {q^2}}V_0(q^2),
\end{eqnarray}
with  $N_1=  {iG_F}V_{ub}/{\sqrt 2}$.
For the sake of convenience, we use
\begin{eqnarray}
 i{\cal A}^1_{\perp/||}&=&\frac{1}{\sqrt 2}[i{\cal A}^1_{+} \mp i{\cal A}^1_{-}],\nonumber\\
i{\cal A}^1_\perp&=& -i \sqrt{ N_{a_1}} \sqrt{2} N_1
 \frac{\sqrt \lambda A(q^2)}{m_{B}-m_{a_1}},\;\;\;
i{\cal A}^1_{||}= i \sqrt{ N_{a_1}}\sqrt{2} N_{1}    (m_{B}-m_{a_1})V_1(q^2).
\end{eqnarray}

The differential decay width for $\overline B^0\to a_1^+\ell^- \nu_\ell$ is then derived  as
\begin{eqnarray}
 \frac{d\Gamma}{dq^2  }
 &=& \frac{3}{8}\Big[2I_1  -\frac{2}{3} I_2     \Big],
\end{eqnarray}
with the $I_i$ having the form:\begin{eqnarray}
 I_1  &=&   \left[(1+\hat m_l^2) |{\cal A}^1_{0}|^2
 +2 \hat m_l^2  |{\cal A}_t^1|^2\right]   +    \frac{3+\hat m_l^2}{2}    [|{\cal A}^1_{\perp}|^2+|{\cal A}^1_{||}|^2 ],\nonumber\\
 I_2   &=& -\beta_l         |{\cal A}^1_{0}|^2     +
 \frac{1}{2}\beta_l     (|{\cal A}^1_{\perp}|^2+|{\cal A}^1_{||}|^2).
 \label{eq:simplified_angularCoefficients}
\end{eqnarray}

\begin{figure}\begin{center}
\includegraphics[scale=0.6]{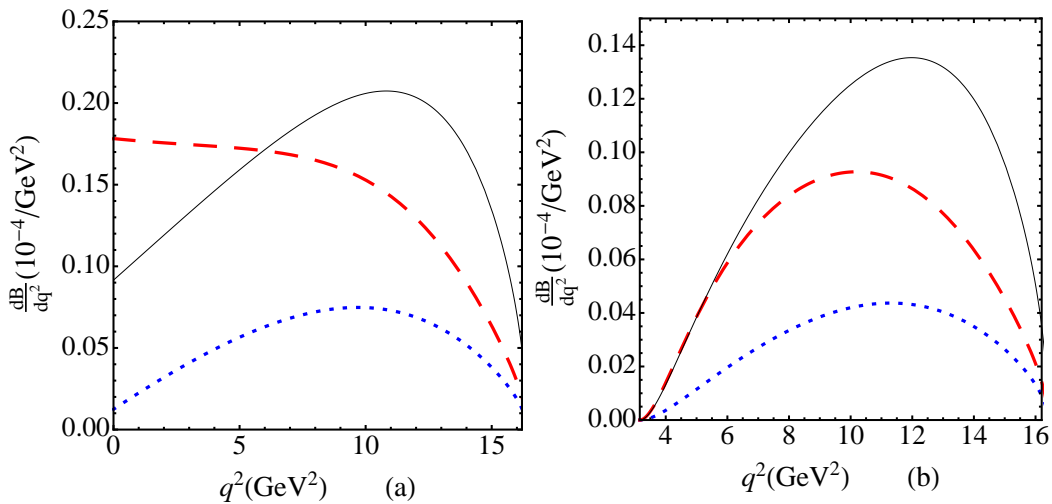}
\caption{Differential branching fractions $d{\cal B}/dq^2$ (in units of $10^{-4}/{\rm GeV}^2$) for the decay $\overline B^0\to a_1^+\ell^-\bar\nu$.  The left panel corresponds to  $\ell=(e,\mu)$ and the right panel corresponds to $\ell=\tau$.  The dotted, dashed and  solid curves are obtained using the form factors calculated in LFQM, LCSR and PQCD approach.   } \label{fig:dBdq2}
\end{center}
\end{figure}

With the above formulas at hand, we present our results for  differential branching ratios $d{\cal B}/dq^2$ (in units of $10^{-4}/{\rm GeV}^2$) in Fig.~\ref{fig:dBdq2}. The left and right  panel corresponds to  $\ell=(e,\mu)$ and $\ell=\tau$, respectively.  The dotted, dashed and  solid curves are obtained using the  LFQM~\cite{Cheng:2003sm}, LCSR~\cite{Yang:2008xw} and PQCD~\cite{Li:2009tx} form factors.
The other input parameters are taken from Particle Data Group (PDG)~\cite{Agashe:2014kda} as follows:
\begin{eqnarray}
 m_B=5.28 {\rm GeV},\quad \tau_{B^0}=1.52\times 10^{-12}{\rm s}, \quad m_{a_1}=1.23 {\rm GeV}, \nonumber\\
m_{e}=0.511  {\rm MeV}, \quad m_{\mu}=0.106 \rm{GeV}, \quad m_{\tau}=1.78 {\rm GeV},\nonumber\\
G_F=1.166\times 10^{-5} {\rm GeV}^{-2}, \quad |V_{ub}|=(3.28\pm0.29)\times 10^{-3}.
\end{eqnarray}
As for the $|V_{ub}|$,  we have quoted the value extracted from the exclusive $B\to \pi\ell\bar\nu_{\ell}$ for self-consistent, see Refs.~\cite{Wang:2014sba,Ricciardi:2014aya} for  discussions on the so-called $|V_{ub}|$ puzzle.

Integrating over the $q^2$, one obtains the longitudinal and transverse contributions to branching fractions
of $\overline B^0\to a_1^+ \ell^-\bar\nu$ decays, and our  results are given in Table~\ref{tab:br_Ba1}.
Uncertainties shown in the table  arise from the ones in the $B\to a_1$ form factors.
We can see from this table that  branching fractions for the $ \overline B^0\to a_1^+ \ell^-\bar\nu$ are of the order $10^{-4}$.
These  values  are comparable to the data by Belle collaboration  on  branching fractions for the semileptonic $B$ decays into a vector meson~\cite{Sibidanov:2013rkk}:
\begin{eqnarray}
 {\cal B}( B^- \to \rho^0 \ell^-\bar\nu) &=& (1.83\pm0.10\pm0.10)\times 10^{-4}, \\
 {\cal B}(\overline B^0 \to \rho^+ \ell^-\bar\nu) &=& (3.22\pm0.27\pm0.24)\times 10^{-4}, \\
 {\cal B}(B^- \to \omega \ell^-\bar\nu) &=& (1.07\pm0.16\pm0.07)\times 10^{-4}.
\end{eqnarray}
Babar collaboration~\cite{Lees:2012vv} also gives similar results for the $B^- \to \omega \ell^-\bar\nu$:
\begin{eqnarray}
 {\cal B}(B^- \to \omega \ell^-\bar\nu) &=& (1.19 \pm 0.16\pm 0.09)\times 10^{-4}.
\end{eqnarray}
Currently, there is no experimental analysis of the $\overline B^0\to a_1(1260)^+\ell^-\bar\nu$, but the two B factories at KEK and SLAC have accumulated about $10^9$ events of $B^0$ and $B^\pm$.
The branching fractions  ${\cal O}(10^{-4})$ correspond  to about $10^5$ events  for the signal.  The above  estimate may be affected by the detector efficiency, but an experimental search would very presumably lead to the observation of this decay mode.
In addition, the sizable branching fractions as shown  in Table~\ref{tab:br_Ba1} also indicate a promising prospect at the ongoing  LHC experiment~\cite{Bediaga:2012py}, the forthcoming   Super-KEKB factory~\cite{Aushev:2010bq} and the under-design   Circular Electron-Positron Collider (CEPC)~\cite{CEPC_preCDR}.

 \begin{table}
 \caption{Integrated branching ratios for the $\overline B^0 \to a_1^+(1260)\ell^-\bar\nu$ decays (in  units of $10^{-4}$) with the form factors from the LFQM~\cite{Cheng:2003sm}, LCSR~\cite{Yang:2008xw} and PQCD~\cite{Li:2009tx}. }
 \label{tab:br_Ba1}
 \begin{center}
 \begin{tabular}{|c|cccc|c|cccc|}
 \hline\hline
     $\ell=(e,\mu)$ & ${\cal B}_{\rm{L}}$  &${\cal B}_{\rm{T}}$    &${\cal B}_{\rm{total}}$  &${\cal B}_{\rm{L}}/{\cal B}_{\rm{T}}$   &   $\ell=\tau $ &  ${\cal B}_{\rm{L}}$  &${\cal B}_{\rm{T}}$    &${\cal B}_{\rm{total}}$  &${\cal B}_{\rm{L}}/{\cal B}_{\rm{T}}$\\
 \hline
 \ \ \        LFQM  &$0.22$                     &$0.65$                  &$0.87$                  &$0.33$
             &LFQM  &$0.09$                     &$0.28$                  &$0.38$                  &$0.33$ \\
 \hline
 \ \ \        LCSR  &$0.74_{-0.26}^{+0.31}$     &$1.60_{-0.54}^{+0.66}$  &$2.34_{-0.80}^{+0.97}$  &$0.46$
             &LCSR  &$0.16_{-0.05}^{+0.06}$     &$0.67_{-0.23}^{+0.28}$  &$0.83_{-0.28}^{+0.34}$  &$0.24$\\
 \hline
 \ \ \        PQCD &$1.42_{-0.72}^{+0.99}$      &$1.21_{-0.50}^{+0.71}$  &$2.63_{-1.21}^{+1.71}$  &$1.17$
             &PQCD &$0.61_{-0.31}^{+0.42}$      &$0.58_{-0.24}^{+0.34}$  &$1.19_{-0.54}^{+0.77}$  &$1.05$  \\
 \hline
 \end{tabular}
 \end{center}
 \end{table}

\subsection{Nonleptonic $B$ decays into $a_1$}

Since our main goal in this work is to investigate   the internal structure of the $a_1(1420)$, we will focus on the decay modes  which  can be handled under the factorization approach. These decay modes are typically dominated by  tree operators with effective Hamiltonian
\begin{eqnarray}
 {\cal H}_{\rm eff} &=& \frac{G_{F}}{\sqrt{2}}V_{ub} V_{ud}^{*}
     \Bigg\{
     C_{1} [{\bar{u}}_{\alpha}\gamma^\mu (1-\gamma_5)b_{\beta}]
               [{\bar{d}}_{\beta}\gamma_\mu (1-\gamma_5) u_{\alpha}]
  +  C_{2} [{\bar{u}}_{\alpha}\gamma^\mu (1-\gamma_5)b_{\alpha}]
               [{\bar{d}}_{\beta} \gamma_\mu (1-\gamma_5)u_{\beta} ] \Bigg\},
 \label{eq:hamiltonian01}
 \end{eqnarray}
 where $C_1$ and $C_2$ are the Wilson coefficients. The  $\alpha$ and $\beta$ are the color indices. $V_{ub}, V_{ud}$ are the CKM matrix elements.   The up type quark can also be replaced by the charm quark.

With the definitions of  decay constants,
\begin{eqnarray}
  \langle a_1(p,\epsilon)|\bar d \gamma_\mu \gamma_5 u|0\rangle
   &= & if_{a_1}  m_{a_1} \,  \epsilon^{*}_\mu,\;\;\;
  \langle J/\psi(p,\epsilon)|\bar c\gamma_\mu c|0\rangle
   =   f_{J/\psi} m_{J/\psi} \,  \epsilon^{*}_\mu,\nonumber\\
  \langle \pi^-(p)|\bar d \gamma_\mu \gamma_5 u|0\rangle
   &= & -if_{\pi}  p_\mu,
 \end{eqnarray}
we expect the factorization formula to have the form:
\begin{eqnarray}
 i{\cal A}(\overline B^0\to D^+ a_1^-) &=&  (-i)^2\frac{G_F}{\sqrt 2}   V_{cb}V_{ud}^* a_1f_{a_1}
 F_1^{B\to D}(m_{a_1}^2)  m_B^2 \sqrt{\lambda(1, r_D^2, r_{a_1}^2)},\\
 i {\cal A}(\overline B^0\to \pi^+ a_1^-) &=&   (-i)^2\frac{G_F}{\sqrt 2} V_{ub}V_{ud}^* a_1  f_{a_1}
 F_1^{B\to \pi}(m_{a_1}^2) m_B^2 \sqrt{\lambda(1, r_\pi^2, r_{a_1}^2)},\\
 i {\cal A}(\overline B^0\to \pi^- a_1^+) &=&   (-i)^2\frac{G_F}{\sqrt 2}  V_{ub}V_{ud}^* a_1 f_{\pi}
 V_0^{B\to a_1}(m_{\pi}^2) m_B^2 \sqrt{\lambda(1, r_\pi^2, r_{a_1}^2)} ,\end{eqnarray}
 \begin{eqnarray}
i {\cal A}_L(\overline B^0\to D^{*+} a_1^-)&=& \frac{ (-i)^3 G_F}{\sqrt 2} V_{cb}V_{ud}^* a_1   f_{a_1}  m_{B}^2 \frac{ 1}{2 r_{D^*}} \nonumber\\
 && \times \left[(1-r_{D^*}^2-r_{a_1}^2)(1+r_{D^*})A_1^{B\to D^*}(m_{a_1}^2)- \frac{{\lambda(1,r_{D^*}^2, r_{a_1}^2) }}{1+r_{D^*}}A_2^{B\to D^*}(m_{a_1}^2)\right],\nonumber\\
i {\cal A}_N(\overline B^0\to D^{*+} a_1^-)&=& \frac{ (-i)^3 G_F}{\sqrt 2} V_{cb}V_{ud}^* a_1  f_{a_1} m_B^2 (1+r_{D^*})   r_{a_1} A_1^{B\to D^*}(m_{a_1}^2),\nonumber\\
i {\cal A}_T(\overline B^0\to D^{*+} a_1^-)&=&
 \frac{-i G_F}{\sqrt 2}
 V_{cb}V_{ud}^* a_1   f_{a_1} r_{a_1}  m_B^2\frac{\sqrt {\lambda(1,r_{a_1}^2,r_{D^*}^2)}}{(1+r_{D^*})}
V^{B\to D^*}(m_{a_1}^2),
\end{eqnarray}

\begin{eqnarray}
i {\cal A}_L(B^-\to a_1^-J/\psi)&=& \frac{ (-i)^3G_F}{\sqrt 2}V_{cb}V_{cd}^* a_2   f_{J/\psi}m_{B}^2\frac{ 1}{2 r_{a_1}} \nonumber\\
&&\times \left[(1-r_{J/\psi}^2-r_{a_1}^2)(1-r_{a_1})V_1^{B\to a_1}(m_{J/\psi}^2)-\frac{{\lambda(1,r_{J/\psi}^2, r_{a_1}^2)}}{1-r_{a_1}}V_2^{B\to a_1}(m_{J/\psi}^2)\right],\nonumber\\
 i{\cal A}_N(B^-\to a_1^-J/\psi)&=&
 \frac{ (-i)^3G_F}{\sqrt 2}
 V_{cb}V_{cd}^* a_2  f_{J/\psi} m_{B}^2 r_{J/\psi} (1-r_{a_1})  V_1^{B\to a_1}(m_{J/\psi}^2),\nonumber\\
i {\cal A}_T(B^-\to a_1^- J/\psi)&=& \frac{-i G_F}{\sqrt 2}
 V_{cb}V_{cd}^* a_2   f_{J/\psi}m_{B}^2  r_{J/\psi} \frac{\sqrt {\lambda(1, r_{J/\psi}^2,r_{a_1}^2)}}{1-r_{a_1}} A^{B\to a_1}(m_{J/\psi}^2),
\end{eqnarray}
with $a_1= C_2+C_1/N_c$ and $a_2= C_1+C_2/N_c(N_c=3)$.
In the above, the amplitude for the  $B\to J/\psi a_1$ has been decomposed according to
the Lorentz structures
\begin{eqnarray}
 {\cal A}&=& {\cal A}_{L} +\epsilon_{J/\psi}^*(T)\cdot \epsilon_{a_1}^*(T)
 {\cal A}_N + i {\cal
 A}_T\epsilon_{\alpha\beta\gamma\rho}\epsilon^{*\alpha}_{J/\psi}\epsilon^{*\beta}_{a_1}
 \frac{2p_{J/\psi}^{\gamma}p_{a_1}^\rho}
 {\sqrt {\lambda(m_{B}^2, m_{J/\psi}^2,m_{a_1}^2)}}.
\end{eqnarray}
The partial decay width of the $B\to a_1P$, where $P$ denotes a pseudoscalar meson, is given as
\begin{eqnarray}
 \Gamma(B\to a_1
P)&=& \frac{|\vec p|}{8\pi m_{B}^2} \left|{\cal
 A}(B\to a_1
P)\right|^2,
\end{eqnarray}
with $|\vec p|$ being the three-momentum of the $a_1$ in the $B$
meson rest frame.  For the $B\to a_1 V$, the partial decay
width is the summation of   three polarizations
\begin{eqnarray}
 \Gamma(B\to a_1 V)&=& \frac{|\vec p|}{8\pi m_{B}^2} \left(\left|{\cal
 A}_0(B\to a_1V)\right|^2+2\left|{\cal
 A}_{N}(B\to a_1V)\right|^2+2\left|{\cal
 A}_{T}(B\to a_1V)\right|^2\right).
\end{eqnarray}

We use the LFQM results~\cite{Cheng:2003sm} for all transition form factors and other inputs are given as~\cite{Agashe:2014kda} 
\begin{eqnarray}
\tau_{B^-}= (1.638\times 10^{-12}) s,\;\;\;
\tau_{B_s}= (1.511\times 10^{-12}) s\\
|V_{cb}|=41.1\times10^{-3},\quad|V_{ud}|=0.974,\quad|V_{cd}|=0.225,
\end{eqnarray}
The  $f_\pi$ and $f_{J/\psi}$ can be extracted from the $\pi^-\to \ell^-\bar\nu$ and $J/\psi\to \ell^+\ell^-$ data~\cite{Agashe:2014kda}:
\begin{eqnarray}
f_\pi= 130.4{\rm MeV}, \;\;\; f_{J/\psi}= (416.3\pm 5.3){\rm MeV}.
\end{eqnarray}
We use   QCD sum rules results for the $f_{a_1}$~\cite{Yang:2007zt}
\begin{eqnarray}
 f_{a_1}= (238\pm 10){\rm MeV}.
\end{eqnarray}
The Wilson coefficient $a_1$ is used as~\cite{hep-ph/9512380}
\begin{eqnarray}
 a_1=1.07,
\end{eqnarray}
while under the same factorization hypothesis the $a_2$ is extracted from the $B\to J/\psi K^*$ data~\cite{Agashe:2014kda} :
\begin{eqnarray}
 a_2= (0.234\pm 0.006).
\end{eqnarray}
Then theoretical  results  for branching ratios are given as
\begin{eqnarray}
 {\cal B}(\overline B^0\to D^+ a_1^-) &=& (1.3\pm 0.1)\%, \label{eq:br_B_D_a1}\\
 {\cal B}(\overline B^0\to \pi^+ a_1^-)&=& (1.9\pm0.2)\times 10^{-5}, \\
 {\cal B}(\overline B^0\to D^{*+} a_1^-)&=& (1.6\pm0.2)\%,
\end{eqnarray}
where the errors come from the one in the $f_{a_1}$. For   decay modes induced by the $B\to a_1$ transition, we have
\begin{equation}
{\cal B}(\overline{B}^{0}\to\pi^{-}a_{1}^{+})=\begin{cases}
0.13\times10^{-5}, & {\rm LFQM}\\
(0.70_{-0.22}^{+0.25})\times10^{-5}, & {\rm LCSR}\\
(0.89_{-0.48}^{+0.68})\times10^{-5}, & {\rm PQCD}
\end{cases} \label{eq:br_B_a1_pi}
\end{equation}
\begin{equation}
{\cal B}(B^{-}\to a_{1}^{-}J/\psi)=\begin{cases}
3.6\times10^{-5} ,& {\rm LFQM}\\
(7.5_{-2.5}^{+3.1})\times10^{-5}, & {\rm LCSR}\\
(9.8_{-4.4}^{+6.3})\times10^{-5}, & {\rm PQCD}\label{eq:br_Bbar_a1_pi}
\end{cases}
\end{equation}
where the errors arise from those in form factors. 

Babar~\cite{Aubert:2006dd} and Belle~\cite{Dalseno:2012hp} collaborations have reported the observation of $B^0\to a_1^\pm\pi^\mp$ and their results for branching fractions are given as
\begin{eqnarray}
 {\cal B}(\overline B^0\to \pi^\pm a_1^\mp){\cal B}(a_1^\pm\to\pi^\pm \pi^\mp\pi^\pm)=
\left\{\begin{array}{ccc}
 (16.6\pm1.9\pm1.5)\times 10^{-6}, \;\;\;{\rm Babar}   \\
  (11.1\pm   1.0\pm 1.4)\times 10^{-6}.\;\;\;{\rm Belle}
\end{array}
\right.
\end{eqnarray}
The above results  have been  averaged by PDG as~\cite{Agashe:2014kda}
\begin{eqnarray}
 {\cal B}(\overline B^0\to \pi^\pm a_1^\mp)= (2.6\pm 0.5)\times 10^{-5}.
\end{eqnarray}
As we can see, the averaged data is consistent with our theoretical results in Eqs.~(\ref{eq:br_B_a1_pi}) and (\ref{eq:br_Bbar_a1_pi}).

We also predict the branching ratios for  $\overline B_s^0\to D_s^+ a_1^-$ and $\overline B_s^0\to D_s^{*+} a_1^-$:
\begin{eqnarray}
 {\cal B}(\overline B_s^0\to D_s^+ a_1^-) &=& (1.3\pm0.1)\%.  \\
 {\cal B}(\overline B_s^0\to D_s^{*+} a_1^-) &=&(1.7\pm0.2)\%.  \label{eq:br_Bs_Ds_a1}
\end{eqnarray}
The numerical  results given in Eqs.~(\ref{eq:br_B_D_a1}-\ref{eq:br_Bs_Ds_a1}) indicate that there is a promising prospect to  study  these decays by the LHCb, Babar,  and Belle collaborations,  and on  the forthcoming  Super-KEKB factory and the CEPC.

\section{$D\to a_1$ decays}
\label{sec:Da1}

\begin{figure}\begin{center}
\includegraphics[scale=0.5]{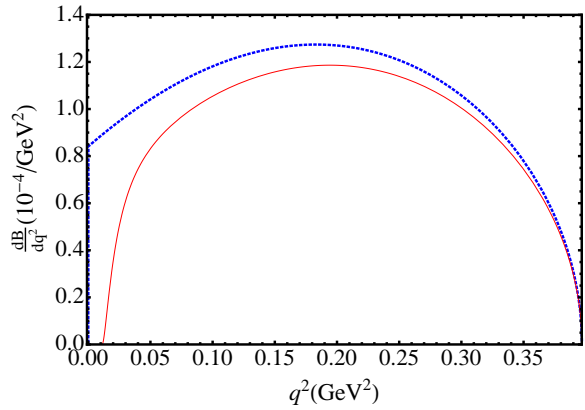}
\caption{Differential branching fractions $d{\cal B}/dq^2$ (in units of $10^{-4}/{\rm GeV}^2$) for the decay $D^0\to a_1^-\ell^+\nu$.  The dotted and solid curve corresponds to  $\ell=e$  and $\ell=\mu$, respectively. The differences between the two curves arise from the lepton masses and can reach  about $10\%$.     } \label{fig:dBdq2D_a1}
\end{center}
\end{figure}

By replacing the corresponding form factors and CKM matrix elements, the analysis of $B$ decays   in the last section can be straightforwardly  generalized  to the $D\to a_1$ decays.
The  $D\to a_1$ form factors are only available in LFQM~\cite{Cheng:2003sm} and we summarize these results  in Table~\ref{Tab:formFactor_D_a1}.  We will use  other input parameters  as follows~\cite{Agashe:2014kda}:
\begin{eqnarray}
m_{D^0}=1.8648 {\rm GeV},\;\; |V_{cd}|=0.225, \;\; \tau_{D^0}=0.410\times 10^{-12}{\rm s},\;\; \tau_{D_s}=0.500\times 10^{-12}{\rm s}
 \end{eqnarray}
Our results for  the differential branching ratios of the semileptonic $D^0\to a_1^-\ell^+\nu$ are given in Fig.~\ref{fig:dBdq2D_a1}, and  the integrated branching fractions are presented in Table~\ref{tab:br_Da1}. In Fig.~\ref{fig:dBdq2D_a1}, the dotted and solid curve corresponds to  $\ell=e$  and $\ell=\mu$, respectively. The differences in the two curves arise from the lepton masses and can reach  about $10\%$.

Recently, based on the $2.9fb^{-1}$ data of electron-positron annihilation data collected at a center-of-mass energy of $\sqrt{s}=3.773$ GeV, BES-III collaboration has searched for  the $D^+\to \omega\ell^+\nu$ decay~\cite{Ablikim:2015gyp} and the branching fraction is measured
\begin{eqnarray}
{\cal B}(D^+\to \omega \ell^+\nu)&=& (1.63\pm0.11\pm 0.08)\times 10^{-3}.
\end{eqnarray}
In this procedure, the $\omega$ meson is reconstructed by three pions, and it is interesting to notice that  the neutral $a_1(1260)$ should also be reconstructed by the same final state.  Extending the analysis in Ref.~\cite{Ablikim:2015gyp} to higher mass region at round $1.23$ GeV may discover the $D^+\to a_1^0\ell^+\nu$ transition.  Actually,  BES-III have collected about $10^7$ events of  $D-\bar D$. The $10^{-4}$ branching fractions correspond to about $10^3$ events for the $D\to a_1\ell^+\nu$, which might be observed in the future analysis.

\begin{table}
\caption{The $D\to a_1$ form factors calculated in the covariant  LFQM~\cite{Cheng:2003sm}.  }
 \label{Tab:formFactor_D_a1}
 \begin{center}
 \begin{tabular}{|c|c|c|c|}
\hline \hline
         $F$   &$F(0)$   &$a$   &$b$            \\
\hline
\hline
         $A^{D\to a_1}$          &$0.20$       &$0.98$     &$0.20$       \\
\hline   $V_0^{D\to a_1}$        &$0.31$       &$0.85$     &$0.49$       \\
\hline   $V_1^{D\to a_1}$        &$1.54$       &$-0.05$    &$0.05$       \\
\hline   $V_2^{D\to a_1}$        &$0.06$       &$0.12$     &$0.10$       \\
\hline
\end{tabular}
\end{center}
\end{table}

 \begin{table}
 \caption{Integrated branching ratios for the $D^0\to a_1^-(1260)\ell^+\nu$ decays (in  units of $10^{-4}$). }
 \label{tab:br_Da1}
 \begin{center}
 \begin{tabular}{|c|cccc|c|cccc|}
 \hline\hline
      & ${\cal B}_{\rm{L}}$  &${\cal B}_{\rm T}$    &${\cal B}_{\rm{total}}$  &${\cal B}_{\rm{L}}/{\cal B}_{\rm{T}}$    \\
 \hline
 \ \ \    $\ell=e$     &$0.21$     &$0.20$  &$0.41$  &$1.05$   \\
 \ \ \     $\ell=\mu$      &$0.18$     &$0.18$  &$0.36$  &$1.00$ \\
 \hline
 \end{tabular}
 \end{center}
 \end{table}

We can also study the nonleptonic $D/D_s$ decays into $a_1(1260)$  with the factorization amplitudes:
\begin{eqnarray}
i {\cal A}(D^0\to a_1^+\pi^- ) &=&  (-i)^2\frac{G_F}{\sqrt 2} V_{cd}^*V_{ud} a_1  f_{a_1}
 F_1^{D\to \pi}(m_{a_1}^2)  \sqrt{\lambda(m_D^2, m_{\pi}^2, m_{a_1}^2)},\\
i {\cal A}(D^0\to \pi^+ a_1^-) &=&  (-i)^2\frac{G_F}{\sqrt 2} V_{cd}^*V_{ud} a_1 f_{\pi}
 V_0^{D\to a_1}(m_{\pi}^2)   \sqrt{\lambda(m_D^2, m_{\pi}^2, m_{a_1}^2)},\\
i {\cal A}(D_s^+ \to a_1^+K^0 ) &=& (-i)^2\frac{G_F}{\sqrt 2}V_{cd}^*V_{ud} a_1  f_{a_1}
 F_1^{D_s\to K}(m_{a_1}^2)   \sqrt{\lambda(m_{D_s}^2, m_{K}^2, m_{a_1}^2)}.
 \end{eqnarray}
Our theoretical results are given as:
\begin{eqnarray}
 {\cal B}(D^0\to a_1^+ \pi^-) &=& (4.1\pm0.4)\times 10^{-3}, \label{eq:Da1ppim}\\
 {\cal B}(D^0\to \pi^+ a_1^-)&=&  8.1\times 10^{-5}, \label{eq:Dpipa1m}\\
 {\cal B}(D_s^+\to a_1^+ K^0)&=&  (2.3\pm0.2)\times 10^{-3},\label{eq:Dsa1K}
\end{eqnarray}
where the errors arise from those in the decay constant $f_{a_1}$.

The FOCUS collaboration has measured the branching fraction for the $D^0\to a_1^\pm\pi^\mp$~\cite{Link:2007fi}:
\begin{eqnarray}
{\cal B}(D^0\to a_1^\pm\pi^\mp )= (4.47\pm0.32)\times 10^{-3},
\end{eqnarray}
which is consistent with our theoretical results in Eqs.~(\ref{eq:Da1ppim}-\ref{eq:Dpipa1m}). The
LHCb collaboration makes use of the $D^0\to a_1^\pm\pi^\mp$  to study CP violation~\cite{Aaij:2013swa}, and it is also feasible to study this mode using the CLEO-c data~\cite{Benton:2013zra}.  The BES-III collaboration  has accumulated about $10^7$ events of the $D^0$ and will collect about  $3fb^{-1}$ data  at the center-of-mass $\sqrt s= 4.17$ GeV to produce the  $D_s^+D_s^-$~\cite{Ablikim:2014cea,Asner:2008nq}. All these data can be used to study the charm decays into the $a_1$.

\section{Conclusions}

Experimental observations of resonance-like states in recent years  have  invoked theoretical research  interest on exotic  hadron spectroscopy. In particular, many of the experimentally established structures  defy the naive quark model  assignment as a  $\bar qq$ or $qqq$ state.  
At the low-energy, the  $a_1(1420)$   with $I^G(J^{PC})= 1^-(1^{++})$  observed in the $\pi^+ f_0(980)$ final state  in the $\pi^-p\to \pi^+\pi^-\pi^- p$   process  by  COMPASS collaboration seems unlikely to be an ordinary  $\bar qq$ mesonic state. Available theoretical explanations include tetraquark or rescattering effects due to $a_1(1260)$ decays.  If the $a_1(1420)$ were induced by   rescattering effects,  its production rates are completely determined by those of the $a_1(1260)$.

In this work, we have  proposed to explore the ratios of branching fractions of  heavy meson weak decays into  the $a_1(1420)$ and $a_1(1260)$, and testing the universality of these ratios  would be a straightforward way to validate/invalidate the rescattering explanation.
The decay modes include in the charm sector  the $D^0\to  a_1^-\ell^+\nu$ and  $D\to \pi^\pm a_1^\mp$, and  in the bottom sector  $B\to  a_1 \ell \bar\nu$ and  $B\to D  a_1, \pi^\pm a_1^\mp$, and the $B_c\to J/\psi a_1$ and $\Lambda_b\to \Lambda_c a_1$.
We have calculated the branching ratios for   various decays into the $a_1(1260)$.
Other decay modes like $\Lambda_b\to \Lambda_c a_1$, and  $B_c^-\to J/\psi a_1^-$ which has been measured by the LHCb collaboration~\cite{LHCb:2012ag} and CMS collaboration~\cite{Khachatryan:2014nfa}, in agreement with theoretical results based on  the form factors~\cite{Wang:2008xt,Likhoded:2009ib}, are also of helpful in this aspect.  

Our results have indicated  that there is a promising prospect to  study  these decays on experiments including BES-III, LHCb, Babar, Belle and CLEO-c,  the forthcoming  Super-KEKB factory and the under-design Circular Electron-Positron Collider. Experimental analyses in future will very probably  lead to a deeper understanding of the nature of the $a_1(1420)$.

\section*{Acknowledgements}
The authors are grateful to Prof. Hai-bo Li,  Dr. Jian-Ping Dai,  and Yong Huang  for  enlightening discussions. This work was supported in part  by National  Natural  Science Foundation of China under Grant  No.11575110,  Natural  Science Foundation of Shanghai under Grant  No. 11DZ2260700, 15DZ2272100 and No. 15ZR1423100,  by the Open Project Program of State Key Laboratory of Theoretical Physics, Institute of Theoretical Physics, Chinese  Academy of Sciences, China (No.Y5KF111CJ1), and  by   Scientific Research Foundation for   Returned Overseas Chinese Scholars, State Education Ministry.



\end{document}